\documentclass[a4paper,conference]{IEEEtran}
\IEEEoverridecommandlockouts
\usepackage{cite}
\usepackage{amsmath,amssymb,amsfonts}
\usepackage{bm}
\usepackage{graphicx}
\usepackage{graphics}
\usepackage{algorithm}
\usepackage{algorithmic}
\usepackage[algo2e,linesnumbered,ruled]{algorithm2e}
\usepackage{multirow}
\usepackage{array}
\usepackage{textcomp}
\usepackage{xcolor}
\def\BibTeX{{\rm B\kern-.05em{\sc i\kern-.025em b}\kern-.08em
    T\kern-.1667em\lower.7ex\hbox{E}\kern-.125emX}}

\newtheorem{definition}{Definition}
\newtheorem{thm}{Theorem}

\SetKwProg{Fn}{Function}{}{end}

\SetAlgoSkip{}

\newcolumntype{M}[1]{>{\centering\arraybackslash}m{#1}}

\DeclareSymbolFont{symbolsC}{U}{txsyc}{m}{n}
\DeclareMathSymbol{\notniFromTxfonts}{\mathrel}{symbolsC}{61}

\makeatletter
\newcommand*\titleheader[1]{\gdef\@titleheader{#1}}
\AtBeginDocument{%
	\let\st@red@title\@title
	\def\@title{%
		\vskip-1.1em\bgroup\normalfont\footnotesize\centering\@titleheader\par\egroup\vskip.4em\st@red@title\vskip-.1em}
}
\makeatother

\title{Highly Efficient Indexing Scheme for $k$-Dominant Skyline Processing over Uncertain Data Streams}

\titleheader{This is the accepted version of the work. The final version will be published in 2021 IEEE WOCC, October 7-8, Taipei, Taiwan}

\author{\IEEEauthorblockN{Chuan-Chi Lai\IEEEauthorrefmark{1}, Hsuan-Yu Lin\IEEEauthorrefmark{2}, and Chuan-Ming Liu\IEEEauthorrefmark{2}}
	\IEEEauthorblockA{\IEEEauthorrefmark{1}Deptartment of Information Engineering and Computer Science, Feng Chia University, Taichung, Taiwan\\
		\IEEEauthorrefmark{2}Deptartment of Computer Science and Information Engineering, National Taipei University of Technology, Taipei, Taiwan\protect\\
		\IEEEauthorrefmark{1}Email: chuanclai@fcu.edu.tw
	}%
}

\begin{document}


\maketitle

\begin{abstract}
Skyline is widely used in reality to solve multi-criteria problems, such as environmental monitoring and business decision-making. When a data is not worse than another data on all criteria and is better than another data at least one criterion, the data is said to dominate another data. When a data item is not dominated by any other data item, this data is said to be a member of the skyline. However, as the number of criteria increases, the possibility that a data dominates another data decreases, resulting in too many members of the skyline set. To solve this kind of problem, the concept of the $k$-dominant skyline was proposed, which reduces the number of skyline members by relaxing the limit. The uncertainty of the data makes each data have a probability of appearing, so each data has the probability of becoming a member of the $k$-dominant skyline. When a new data item is added, the probability of other data becoming members of the $k$-dominant skyline may change. How to quickly update the $k$-dominant skyline for real-time applications is a serious problem. This paper proposes an effective method, Middle Indexing (MI), which filters out a large amount of irrelevant data in the uncertain data stream by sorting data specifically, so as to improve the efficiency of updating the $k$-dominant skyline. Experiments show that the proposed MI outperforms the existing method by approximately 13\% in terms of computation time.
\end{abstract}


\section{Introduction}\label{sec:introduction}
Nowadays, \emph{Skyline} is widely used in reality. It is used to solve multi-criteria decision problems. Users can search through skyline to find the result that best meets user needs. A variant of skyline,
$k$-dominant skyline query, has been widely studied and many efficient $k$-dominant skyline query methods have been proposed. For example: using a Bitmap Index (BI) to process skyline~\cite{tan2001efficient} or using the parallel computing architecture, MapReduce, to speed up the computing speed of the skyline query~\cite{tian2014efficient}~\cite{zhang2015efficient}.

Most of the existing methods~\cite{borzsony2001skyline}~\cite{zhang2011adapting}
~\cite{huang2018efficient} are based on the certain data model, but not all data is certain/deterministic data in practical applications. For example, the data sent by the sensor may be uncertain due to the aging of the sensor.
The way of calculating the $k$-dominant skyline in an environment with uncertain data~\cite{li2019parallel} is completely different from the way of considering certain data~\cite{chan2006finding}
~\cite{siddique2009k}
~\cite{lee2017efficient}. Even though some related works~\cite{8372693}~\cite{8731646}~\cite{9348055} started considering spatial query processing with edge computing in recent years, the $k$-dominant skyline query method based on edge environment considering uncertain data is rarely discussed.
It prompted us to propose a method to quickly calculate and update the probability of $k$-dominant skyline in an uncertain data environment.

In the considered environment with uncertain data, each project has a specific probability of becoming a $k$-dominant skyline. In general, data continues to flow in and out of the system, and the $k$-dominant probability of each item at different points in time may be different. That is, at each point in time, the $k$-dominance probability of each data item must be updated.
Therefore, calculating and updating the probability of the $k$-dominant skyline in an environment of uncertain data flow requires a huge amount of computation. 

This paper proposes an efficient indexing scheme and applies them to the edge computing environment to effectively filter out irrelevant information. Therefore, the system can reduce unnecessary calculations and speed up the calculation time. The contribution of this work is described below.
\begin{itemize}
	\item Very few research discussed the $k$-dominant skyline query processing over uncertain data streams in edge computing environments.
	\item A new indexing scheme, \emph{Middle Indexing} (MI) method, is proposed for effectively pruning irrelevant information so as to improve the performance of driving $k$-dominant skyline over uncertain data streams.
	\item The simulation result indicates that the proposed method significantly improves the performance of computing $k$-dominant skyline in terms of computation time.
\end{itemize}

The rest of paper is organized as follows. 
Section~\ref{sec:problem} presents the preliminary and problem statement.
The proposed approach with algorithms and examples are  explained in Section~\ref{sec:proposed_approach}. In Section~\ref{sec:simulation}, we present the simulation results. Finally, we make the conclusion remarks of this work in Section~\ref{sec:conclusion}.



\section{Preliminary and Problem Statement}
\label{sec:problem}

\subsection{Preliminary}
Data with uncertainty are called uncertain data~\cite{pei2007probabilistic}. This type of data exists in many applications. For example, there may be multiple temperature sensors at the same location, and the temperature measured by each sensor is different, causing uncertainty in the data. This kind of data uncertainty is common in environments such as environmental testing and location services. The definition of uncertain data is as follows:

%
\begin{definition}[\textbf{Uncertain Data}]
	\label{def:DPDM}
	Given a $d$-dimensional space $S=\{s_1,s_2,\dots,s_d\}$, a set of uncertain data $U=\{u_1,u_2,\dots,u_n\}\in S$, and $u_i \cdot s_j$ represents the $j$th dimensional value of $u_i$, where $i=1,2,\dots,d$ and $j=1,2,\dots,n$, the occurrence probability of uncertain data item $u_i$ can be denoted as $\mathbb{P}(u_i)$.
\end{definition} 

Table~\ref{table:uncertain_data_set} shows an example of an uncertain data set that contains 5 data items. Each data item has 4 attributes and a probability value. In this example, the attribute values of data item $u_1$ in 4 dimensions are 10, 3, 4, and 6, respectively. The occurrance probability of $u_1$ is $\mathbb{P}(u_1)=0.2$.

%

\begin{table}[!t]
	\centering
	\footnotesize
	\caption{Example Of An Uncertain Data Set}
	\label{table:uncertain_data_set} 	
	\begin{tabular}{|l|l|l|l|l|l|}
		\hline
		\textbf{Item} & \textbf{Attr1}& \textbf{Attr2} & \textbf{Attr3} & \textbf{Attr4} & \textbf{Probability}\\
		\hline
		$u_1$ & 10 & 3 & 4 & 6 & 0.2 \\ 
		$u_2$ & 9 & 8 & 5 & 9 & 0.4 \\ 
		$u_3$ & 2 & 10 & 4 & 4 & 0.5 \\ 
		$u_4$ & 5 & 2 & 3 & 8 & 0.1 \\ 
		$u_5$ & 7 & 6 & 4 & 6  & 0.8 \\
		\hline
	\end{tabular}
\end{table}

The data stream continuously flows into the system and continuously accumulates huge data. 
Each data is usually time-stamped and will become obsolete data after a period of time. These obsolete data may bring unimportant information, so it must be filtered out outdated data to ensure that it does not affect the correctness of the calculation. Due to the infinite nature of the data stream, it is impossible to calculate all the data. Using the sliding window model can only calculate the data of interest and filter out those data that may affect the accuracy of the calculation, which can greatly reduce the computation time. In this work, we adopts the count-based sliding window and it is defined as follows:


\begin{definition}[\textbf{Count-Based Sliding Window}]
	\label{def:rnn}
	A sliding window at time $t$ is denoted as $SW_t$. The sliding window will have a maximum size $n$, denoted as $|SW_t|_{\max}=n$. The size of sliding window at time $t$ is denote as $|SW_t|$. In any time, $|SW_t|$ will not exceed the maximum size $n$. That is $|SW_t|\leq n, \forall t$. In addition, the sliding window handles the data items in a First-In-First-Out (FIFO) manner.
\end{definition} 

Assumes that $|SW_t|_{\max}=3$ and one new data item comes into the system at each time step. For instance, $u_1$ comes at $t=1$, $u_2$ comes at $t=2$, and so on. In such a scenario, the sliding window will be full if $t\geq 3$ and need to remove the oldest data item before the insertion of new data item. Table~\ref{table:sliding_window} gives a corresponding example to show the change of sliding window from $t=1$ to $t=5$.

To search the $k$-dominant skyline, the system needs to calculate the dominant relations between different uncertain items. 
The definition of dominate is defined as follows:
\begin{definition}[\textbf{Dominate}]
	\label{def:dominate}
	Given two different data items, $u_a ,u_b\in U$. Item $u_a$ dominates item $u_b$, denote as $u_a\prec u_b$, if and only if $u_a \cdot s_j \leq u_b \cdot s_j, \forall j=1,2,\dots,d$, and $u_a \cdot s_{j'}<u_b \cdot s_{j'}, \exists j'\in \{1,2,\dots,d\}$. 	
\end{definition}

\begin{table}[!t]
	\centering
	\footnotesize
	\caption{Example Of A Sliding Window}
	\label{table:sliding_window} 
	\begin{tabular}{|l|l|l|}
		\hline
		\textbf{Time Slot} & \textbf{Sliding Window} & \textbf{Size} \\ \hline
		1	&  $SW_1=\{u_1\}$  &  $|SW_1|=1$    \\ 
		2	&  $SW_2=\{u_1,u_2\}$  &  $|SW_2|=2$    \\ 
		3	&  $SW_3=\{u_1,u_2,u_3\}$  &  $|SW_3|=3$    \\ 
		4	&  $SW_4=\{u_2,u_3,u_4\}$  &  $|SW_4|=3$    \\ 
		5	&  $SW_5=\{u_3,u_4,u_5\}$  &  $|SW_5|=3$    \\ \hline
	\end{tabular}
\end{table}

Consider the example in Table~\ref{table:uncertain_data_set}, $u_4$ is not worse than $u_2$ in all the attributes (or dimensions) and $u_4$ outperforms $u_2$ in at least one attribute, so we can say $u_4$ dominates $u_2$, denoted as $u_4\prec u_2$.

In general, the dominant relations between different data items have transitivity property. This phenomenon is called the \emph{transitivity of domination} and it is defined as follows:
\begin{definition}[\textbf{Transitivity of Domination}]
	\label{def:transitivity}
	Given three different data items, $u_a, u_b, u_c\in U$. If $u_a\prec u_b$ and $u_b\prec u_c$, such that $u_a\prec u_c$. 
\end{definition} 


With Definitions~\ref{def:dominate} and~\ref{def:transitivity}, $k$-dominant can be defined as
\begin{definition}[\textbf{$\bm{k}$-Dominate}]
	\label{def:k_dominance}
	Given two different data items, $u_a, u_b\in U$, $u_a$ $k$-dominats $u_b$, denoted as $u_a\prec_k u_b$, if only if following two conditions hold simultaneously:
	\begin{enumerate}
		\item $u_a\cdot s_j\leq u_b\cdot s_j, \forall s_j\in S',$ where $\exists S'\subseteq S, |S'|\geq k$;
		\item $u_a\cdot s_{j'}<u_b\cdot s_{j'}, \exists s_{j'}\in S$.
	\end{enumerate}
\end{definition} 

See the example in Table~\ref{table:uncertain_data_set}. Data item $u_1$ are not worse than $u_2$ in $\mathsf{Attr2}$, $\mathsf{Attr3}$ and $\mathsf{Attr4}$ attributes and outperforms $u_2$ in at least one attribute, such as attribute $\mathsf{Attr2}$, so we can say $u_1$ 3-dominates $u_2$, denoted as $u_1\prec_3 u_2$.

According to definitions~\ref{def:dominate} and~\ref{def:k_dominance}, we can know that $k$-dominance is a relaxing variant of dominance (also called $d$-dominance) and $k<d$. Howerver, such a relaxation makes $k$-dominance violate the transitivity of domination. Consider the example in Table~\ref{table:uncertain_data_set}, if $k=2$, $u_1\prec_2 u_3$ and $u_3\prec_2 u_1$. This phenomenon is called \emph{Cyclic Dominance} (CD) relationship. Most of the existing skyline query methods are based on the transitivity of dominance, so they cannot be directly applied to the $k$-dominant skyline query. 

With the above assumptions and definitions, $k$-dominant skyline can be defined as follows:
\begin{definition}[\textbf{$\bm{k}$-Dominant Skyline}]
	\label{def:k_dominant_skyline}
	Given a $d$ dimensional space $S$, for data item $u \in S$, none of data items $u'\in S$ can $k$-dominates $u$, such that $u$ will be the $k$-dominant skyline, which can be expressed as
	\begin{equation*}
		U_{k-sky} = \{u|\nexists u'\prec_k u, u'\neq u, u\in S, u'\in S\}.
	\end{equation*}
\end{definition} 

In Table~\ref{table:uncertain_data_set}, none of data items can $k$-dominate $u_3$ and $u_4$, so $\{u_3,u_4\}$ is the $k$-dominant skyline.

In an environment with uncertain data, each data item has a probability of being the $k$-dominant skyline. The probability of data item $u$ being the $k$-dominant skyline is defined as 
\begin{definition}[\textbf{Probability of Being $\bm{k}$-Dominant Skyline}]
	\label{def:probability_of_being_k_dominant_skyline}
	According to the above assumptions and definitions, the probability of data item $u$ being the $k$-dominant skyline can be expressed as
	\begin{equation}\label{eq:obj_dominant_probability}
		\mathbb{P}_{k-sky}(u)=\mathbb{P}(u)\times\prod_{u'\in S,u'\prec_k u}\left(1-\mathbb{P}(u')\right),
	\end{equation}
	where $u, u'\in SW$.
\end{definition} 

\begin{figure}[!t]
	\centering
	\includegraphics[width=\columnwidth]{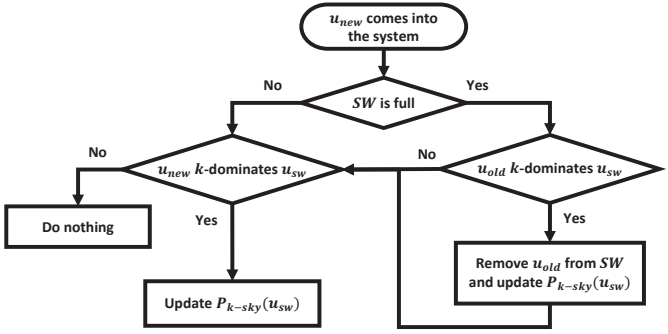}
	\caption{The procedure flowshart of updating $u_{sw}$.}
	\label{fig:fig1}
\end{figure}

If a new data item $u$ comes into the system, denoted as $u_{new}$, the system will derive the probability of $u_{new}$ being the $k$-dominant skyline, $\mathbb{P}_{k-sky}(u_{new})$, using~\eqref{eq:obj_dominant_probability}. Besides, the probability of each data item remaining in the sliding window also needs to be updated. 
The procedure of updating $\mathbb{P}_{k-sky}(u_{sw})$ is depicted in Fig.~\ref{fig:fig1}. If a data item $u$ becomes obsolete, denoted as $u_{old}$, the system will ignore $u_{old}$ and remove it from $SW$. If a data item $u$ is still valid after adding $u_{new}$ into $SW$, denoted as $u_{sw}$, the way of updating the probability of data item $u_{sw}$ being the $k$-dominant skyline will use the following definition:
\begin{definition}[\textbf{$\bm{k}$-Dominant Skyline Probability Update}]
	\label{def:probability_of_being_k_dominant_skyline_update}
	When a new data item $u_{new}$ comes into the system, if the sliding window $SW$ is already full, the system needs to remove an obsolete data item $u_{old}$ from $SW$ in advance. After removing $u_{old}$ from $SW$, the system will update the data item $u_{sw}$ remianing in $SW$ by	
	\begin{align}\label{eq:obj_dominant_probability_update1}
		\mathbb{P}_{k-sky}(u_{sw})=&
		\mathbb{P}_{k-sky}(u_{sw})/(1-\mathbb{P}\left(u_{old})\right),\nonumber\\ 
		& \quad\text{ if } u_{old}\prec_k u_{sw} \wedge |SW|=n.
	\end{align}
	If $SW$ still has free space/slot or the above removal and update have been finished, the system adds $u_{new}$ into $SW$ and derives $\mathbb{P}_{k-sky}(u_{new})$ using~\eqref{eq:obj_dominant_probability}. After that the system update each data item remaining in $SW$ using	
	\begin{align}\label{eq:obj_dominant_probability_update2}
		\mathbb{P}_{k-sky}(u_{sw})=&
		\mathbb{P}_{k-sky}(u_{sw})\times \left(1-\mathbb{P}(u_{new})\right),\nonumber\\
		& \quad\text{ if } u_{new}\prec_k u_{sw}.
	\end{align}
\end{definition} 

\subsection{Problem Statement}
Consider an edge computing environment with uncertain data sources. There are multiple edge computing nodes with adequate computing resources and a main server. All the data comes into edge computing nodes are uncertain data streams. The goal of this work is to quickly calculate and update the $k$-dominant skyline probability of all data items in the sliding window over uncertain data streams. We discuss the performance of $k$-dominant skyline processing using multiple edge computing nodes. In summary, we propose a new method for minimizing the computation time of $k$-dominant skyline processing so as to satisfy the future low-latency data analytic applications.

\section{Proposed Highly Efficient Indexing Scheme}
\label{sec:proposed_approach}
In this section, we will introduce the proposed highly efficient indexing scheme, Middle Indexing (MI), and apply it to distributed edge computing systems. In MI, we propose a new theorem to determine the appropriate thresholds with a sorted indexing table for pruning irrelevant data items, and thus reduce unnecessary comparisons and speed up the overall computation.

\subsection{Data Normalization}
The proposed MI method is designed for breaking the limitation of cross-dimensional comparisons. In order to achieve this goal, the firist stage of MI, \emph{Data Normalization}, will normalize the value of each dimension into the same range, and the normalized items must be sorted in ascending order. An example of a normalized and sorted dataset containing 3 data items is presented in Table~\ref{table:sorted_normalize_exmp}.

\begin{table}[!ht]
	\vspace{-5pt}
	\caption{Example Of A Normalized and Sorted Dataset}
	\label{table:sorted_normalize_exmp}
	\footnotesize 
	\centering	
	\begin{tabular}{|l|l|l|l|}    
		\hline
		Item & Index 0 & Index 1 & Index 2 \\
		\hline
		$u_1$ & 0 & 1 & 1  \\
		$u_2$ & 0 & 0.5 & 1 \\
		$u_3$ & 0 & 0.33 & 0.55  \\
		\hline
	\end{tabular}
	\vspace{-5pt}
\end{table}

\subsection{Threshold Selection}
\normalsize
After the data normalization stage, we can obtain a normalized and sorted dataset $SORTED(U)$ and each data item $u\in U$ will be sorted in ascending order, denoted as $SORTED(u)$. Then, we use a pointer $u_{min}(k)$ to choose an index position and use the value stored in this index position as a threshold $MI_{min}(u,k)$. $u_{min}(k)$ and $MI_{min}(u,k)$ can be respectively expressed as
\begin{align}
	u_{min}(k) &\in \{0,1,...,k-1\}, \label{eq:4}\\
	MI_{min}(u,k) &= SORTED[u_{min}(k)].\label{eq:5}
\end{align}
With~\eqref{eq:4} and~\eqref{eq:5}, we then define another pointer $u_{max}(k)$ and one more threshold $MI_{max}(u,k)$ as 
\begin{align}
	u_{max}(k) &= u_{min}(k) + (d-k), \label{eq:6}\\
	MI_{max}(u,k) &= SORTED[u_{max}(k)]. \label{eq:7}
\end{align}

With the above two pointers and two thresholds, we propose the following theorem:
\begin{thm}
	\label{thm}
	Given a $d$-dimensional space $S$, two data items $p,q\in S$, if $q_{min}(k)=p_{min}(k)$ and $MI_{max}(p,k)<MI_{min}(q,k)$, such that $q$ cannot $k$-dominate $p$.
\end{thm}
\begin{IEEEproof}
	Assume that $q_{min}(k)=p_{min}(k)$, $MI_{max}(p,k)<MI_{min}(q,k)$ and $q$ $(k-1)$-dominates $p$, it means that $q$ is not worse than $p$ in at least $k$ dimensions and $q$ outperforms $p$ in at least one dimension. However, because of $MI_{max}(p,k)<MI_{min}(q,k)$, $q$ has opportunities to outperform $p$ in the dimensions from index $0$ to index $(q_{min}(k)-1)$ and from index $p_{max}(k)$ to index $(d-1)$. In summary, $q$ has at most $1+(q_{min}(k)-1)+(d-1)-p_{max}(k)=k-1$ opportunities to be not worse or better than $p$, which means that $q$ may $(k-1)$-dominate $p$ at most. This contradicts to the given assumption and the proof is done.
\end{IEEEproof}

\subsection{Generation of the Sorted Middle Indexing Table}
The proposed MI method is to sort each item according to $MI_{max}(u,k)$ and $MI_{min}(u,k)$ to filter out items that do not need to be compared, reduce unnecessary calculations and increase the speed of updates. An important aspect of this method is that the index position  $u_{min}(k)$ of each data item $u$ must be the same.
That is, all data items have the same baseline to be sorted.

MI includes two sorting strategies, one is to sort according to the value of $MI_{max}(u,k)$; the other is to sort according to the value of $MI_{min}(u,k)$. The purpose of constucting such an index table is to avoid recalculating and sorting the $MI_{max}(u,k)$ and $MI_{min}(u,k)$ of $u_{sw}$ every time, which can speed up the update of the probabilities of the $k$-dominant skyline items.

Each item that comes in a new stream needs to use $MI_{max}(u,k)$ as the key and insert it into the middle indexing table $MIT_{max}(k)$ in descending order. Similar to the indexing table $MIT_{max}(k)$, another way is to insert each coming new data item into the index table $MIT_{min}(k)$ in ascending order.

\subsection{The Process of Updating $k$-Dominant Skyline Probability}
The procedure flowchart of updating $\mathbb{P}_{k-sky}(u_{sw})$ has been introduced in Fig.~\ref{fig:fig1}. The corresponding algorithm is described as function $\mathsf{MI\_Update}()$ in Algorithm~\ref{alg:mi_update}. The operations from Line~\ref{alg:mi_update:line1} to~\ref{alg:mi_update:line9} implement the update of $\mathbb{P}_{k-sky}(u_{sw})$ in~\eqref{eq:obj_dominant_probability_update1} if the sliding window $SW$ is full. The update of $\mathbb{P}_{k-sky}(u_{sw})$ in~\eqref{eq:obj_dominant_probability_update2} is implemented from Line~\ref{alg:mi_update:line10} to~\ref{alg:mi_update:line16}. Line~\ref{alg:mi_update:line3} and~\ref{alg:mi_update:line11} of Algorithm~\ref{alg:mi_update} use the proposed thresholds in~\eqref{eq:5} and~\eqref{eq:7} to check the conditions $MI_{min}(u_{old},k)>MI_{max}(u_{sw},k)$ and $MI_{min}(u_{new},k)>MI_{max}(u_{sw},k)$, repectively. If $MI_{min}(u_{old},k)>MI_{max}(u_{sw},k)$ holds at Line~\ref{alg:mi_update:line3}, it means that $u_{old}$ cannot $k$-dominates any other items stored in folowing slots of sorted middle indexing table $MIT_{max}(k)$. The system thus does nothing because the following items in the sorted middle indexing table $MIT_{min}(k)$ also cannot $k$-dominates $u_{new}$. Similarly, if $MI_{min}(u_{old},k)>MI_{max}(u_{sw},k)$ holds at Line~\ref{alg:mi_update:line11}, it means $u_{new}$ cannot $k$-dominate any items in $SW$ and the system will do nothing. Otherwise, the system will check whether $u_{new}$ $k$-dominates $u_{sw}$ or not. If yes, the system will update $\mathbb{P}_{k-sky}(u_{sw})$ using~\eqref{eq:obj_dominant_probability_update2} at Line~\ref{alg:mi_update:line14}.
The design of thresholds greatly reduces a lot of unnecessary comparisons. The above benefit has been proved in Theorem~\ref{thm}. 

After updating $\mathbb{P}_{k-sky}(u_{sw})$ of each remaining item $u_{sw}$ in the sliding window $SW$, the systm will calculate the $\mathbb{P}_{k-sky}(u_{new})$ of the new coming item $u_{new}$.
The detailed operations are described as function $\mathsf{MI\_Calculate}()$ in Algorithm~\ref{alg:mi_cal_new}. Line~\ref{alg:mi_cal_new:line2} of Algorithm~\ref{alg:mi_cal_new} uses the proposed thresholds in~\eqref{eq:5} and~\eqref{eq:7} to check the condition $MI_{max}(u_{new},k)<MI_{min}(u_{sw},k)$. If $MI_{max}(u_{new},k)<MI_{min}(u_{sw},k)$ holds, it means that $u_{sw}$ cannot $k$-dominates $u_{new}$. The system thus does nothing because the following items in the sorted middle indexing table $MIT_{min}(k)$ also cannot $k$-dominates $u_{new}$. If $MI_{max}(u_{new},k)<MI_{min}(u_{sw},k)$ does not hold, the system will check whether $u_{sw}$ $k$-dominates $u_{new}$ or not. If yes, the system will calculate/update $\mathbb{P}_{k-sky}(u_{new})$ using~\eqref{eq:obj_dominant_probability} at Line~\ref{alg:mi_cal_new:line5}. 

Finally, the system will add the indexing information of $u_{new}$ to the proposed sorted middle indexing tables $MIT_{max}(k)$ and $MIT_{min}(k)$, which can help the future $k$-dominant skyline processing. Such operations are described as function $\mathsf{MI\_Sort}()$ in Algorithm~\ref{alg:mi_sort}.

We have introduced the key design of our solution with a centralized version algorithm. In fact, MI can be expanded to a distributed version after slight modifications to the edge computing environment. Due to the limited space of this paper, the distributed version algorithm is omitted here.

\begin{algorithm2e}[!t]
	\caption{$\mathsf{MI\_Update}()$}
	\label{alg:mi_update}
	\small
	\SetAlgoLined
	\KwIn{$SW$, $u_{new}$, $u_{old}$, $MIT_{max}(k)$}	
	\KwOut{Updated $SW$}
	\If{$|SW|==n$}{\label{alg:mi_update:line1}
		\ForEach{$e$ in $MIT_{max}(k)$}{
			\tcc{$e$ is $u_{sw}$}
			\uIf{$MI_{min}(u_{old},k)>MI_{max}(e,k)$}{\label{alg:mi_update:line3}
				\textbf{break}\;
			}
			\ElseIf{$u_{old}\prec_k e$}{
				$\mathbb{P}_{k-sky}(e)=		\mathbb{P}_{k-sky}(e)/(1-\mathbb{P}\left(u_{old})\right)$\;
			}
		}
	}\label{alg:mi_update:line9}
	\ForEach{$e$ in $MIT_{max}(k)$}{\label{alg:mi_update:line10}
		\uIf{$MI_{min}(u_{new},k)>MI_{max}(e,k)$}{\label{alg:mi_update:line11}
			\textbf{break}\;
		}
		\ElseIf{$u_{new}\prec_k e$}{
			$\mathbb{P}_{k-sky}(e)=		\mathbb{P}_{k-sky}(e)\times(1-\mathbb{P}\left(u_{new})\right)$\label{alg:mi_update:line14}\;
		}
	}\label{alg:mi_update:line16}
	\Return $SW$\;
\end{algorithm2e}
%

\begin{algorithm2e}[!t]
	\caption{$\mathsf{MI\_Calculate}()$}
	\label{alg:mi_cal_new}
	\small
	\SetAlgoLined
	\KwIn{$SW$, $u_{new}$, $MIT_{min}(k)$}
	\KwOut{$\mathbb{P}_{k-sky}(u_{new})$}
	\ForEach{$e$ in $MIT_{min}(k)$}{
	\uIf{$MI_{max}(u_{new},k)<MI_{min}(e,k)$}{\label{alg:mi_cal_new:line2}
		\textbf{break}\;
	}
	\ElseIf{$e\prec_k u_{new}$}{
		$\mathbb{P}_{k-sky}(u_{new})=		\mathbb{P}_{k-sky}(u_{new})\times(1-\mathbb{P}\left(e)\right)$\label{alg:mi_cal_new:line5}\;
	}
}
	\Return $\mathbb{P}_{k-sky}(u_{new})$\;
\end{algorithm2e}
%

\begin{algorithm2e}[!t]
	\caption{$\mathsf{MI\_Sort}()$}
	\label{alg:mi_sort}
	\small
	\SetAlgoLined
	\KwIn{$u_{new}$}
	\KwOut{$MIT_{max}(k)$, $MIT_{min}(k)$}		
	insert $u_{new}$ to $MIT_{max}(k)$\;
	insert $u_{new}$ to $MIT_{min}(k)$\;
	\Return $MIT_{max}(k)$, $MIT_{min}(k)$\;
\end{algorithm2e}
%


\section{Simulation Results}
\label{sec:simulation}
In this section, we conduct several simulations to verify the performance of the proposed distributed version MI scheme. We compare our approach with the existing method, \textit{PKDS}\cite{li2019parallel}, in an edge computing environment. The simulations are executed on a computer with an Intel Core i7-9700 CPU, 64GB DDR4 RAM, and Windows 10. We deploy 6 virtual machines (5 worker nodes and 1 master node) using Ubuntu 16.04 with Spark 2.4.4 platform. We use python 3.7.4 to implement our simulations. The data in our simulation is the average of 10 runs/results. Table~\ref{table:simulation_settings} shows the default value of each parameter.

\begin{table}[!t]
	\centering
	\caption{Parameter Settings}
	\label{table:simulation_settings} 
	\begin{tabular}{l|l|l}
		\hline
		\textbf{Parameter} & \textbf{Values} & \textbf{Default Value}\\ \hline 
		Data dimensionality & 12 & 12 \\ 
		$k$ & 7, 8, 9, 10, 11 & 11 \\ 
		The size of $SW$ & 300, 400, 500, 600, 700 & 300 \\ 
		The number of data items & 10,000 & 10,000 \\ 
		\hline
	\end{tabular}
	\vspace{-5pt}
\end{table}

\begin{figure}[!t]
	\centering
	\includegraphics[width=.875\columnwidth]{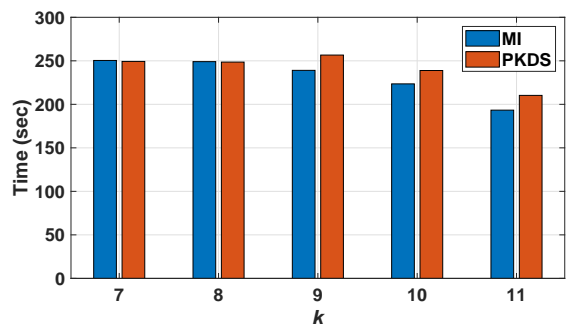}
	\vspace{-5pt}
	\caption{The computation time of different indexing schemes while varying $k$.}
	\label{fig:query_k_value}
	\vspace{-5pt}
\end{figure}

\begin{figure}[!t]
	\centering
	\includegraphics[width=.875\columnwidth]{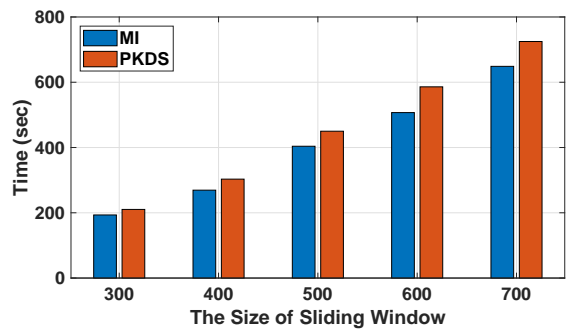}
	\vspace{-5pt}
	\caption{The computation time of different indexing schemes while using diffrent sizes of the sliding windows $|SW|$.}
	\label{fig:SW_size}
	\vspace{-5pt}
\end{figure}

In the simulation, we first observe the impact of $k$ value on the compuation time. The result is shown in Fig.~\ref{fig:query_k_value}. When $k$ becomes larger, the performances of both MI and PKDS in terms of the compuation time are shorter. When $k$ increases, MI will be much better than PKDS by about 8.06\%. The last simulation is to discuss the impact of sliding window size on the computation time. The result is presented in Fig.~\ref{fig:SW_size}. As we can see, if the size of the sliding window increases, the calculation time required becomes the longer. The size of the sliding window ranges from 300 to 700, and MI is about 8\% to 13\% faster than PKDS, and the lead does not change significantly with the size of the sliding window.

\section{Conclusion}
\label{sec:conclusion}
Calculating the $k$-dominant skyline probability of each data item in the uncertain data stream requires a huge amount of computation. The theorem proposed in this paper can effectively and quickly determine the $k$-dominance relationship between two items. We apply this theorem to the derivation of the $k$-dominance skyline. A highly efficient indexing scheme, Middle Indexing, is proposed to effectively filter out a large number of items that do not need to be compared, so that the calculation/update speed is significantly increased. According to the simulation results, MI can increase the performance by about 13\% in terms of calculation time compared with the existing method.

\section*{Acknowledgment}
This research is supported by Ministry of Science and Technology, Taiwan under the Grant No. MOST 109-2221-E-027-095-MY3 and MOST 110-2222-E-035-004-MY2.

\bibliographystyle{IEEEtran}
\bibliography{reference}


\end{document}